# Designing a Potential NASA Fermi Orbit Change

Wayne Yu[a]*, Trevor Williams[b], Russell Carpenter[c]

[a]NASA Goddard, Navigation and Mission Design Branch/595, 8800 Greenbelt Rd., Greenbelt, MD 20771.
wayne.h.yu@nasa.gov
[b]NASA Goddard, Navigation and Mission Design Branch/595, 8800 Greenbelt Rd., Greenbelt, MD 20771.
trevor.w.williams@nasa.gov
[c]NASA Goddard, Space Science Mission Operations/444, 8800 Greenbelt Rd., Greenbelt, MD 20771.
russell.carpenter@nasa.gov
* Corresponding Author

**Abstract**
The Fermi Gamma ray Space Telescope, launched in 2008, has over 16 years of operations providing gamma ray (8 keV to 300 Gev) spectra science observations of cosmic phenomena. It continues to provide invaluable research for the astrophysics community which include the study of pulsars, cosmic rays, gamma ray bursts, and coordination with gravity wave observations for neutron star mergers. The Fermi Earth orbit at a 500 x 512 km altitude is subject to collision warnings due to new constellations deployed near Fermi: currently over 7,000 satellites and growing. This paper presents analysis concerning changing Fermi's orbit and associated operational flight dynamics considerations. The cadence of burns and expected fuel use for a proposed orbit raise scenario is examined, ensuring that Fermi should have sufficient fuel for end-of-life operations. In addition, a Monte Carlo design is presented to capture single maneuver model uncertainty.

**Nomenclature**
Cd: Coefficient of Drag
Cr: Coefficient of Radiation
$\Delta V$: Delta-V
N x M km orbit: N km perigee altitude by M km apogee altitude orbit

**Acronyms/Abbreviations**
AR: Apoapsis Raise Maneuver
AL: Apoapsis Lower Maneuver
CA: Collision Avoidance
CARA: Conjunction Assessment Risk Analysis
DCO: Data Cut-off
FGST: Fermi Gamma-Ray Space Telescope
FDS: Flight Dynamics System for Ground Operations
GMAT: General Mission Analysis Tool Software
GLAST: Gamma-Ray Large Aperture Space Telescope (Original Name of Fermi)
GPS: Global Positioning Satellite
MJ2000: Modified Julian Year 2000 Inertial Frame
MC: Monte Carlo
MP: Maneuver Planning
OD: Orbit Determination
PR: Periapsis Raise Maneuver
PL: Periapsis Raise Lowering
RIC: Radial-Intrack-Crosstrack Local Coordinate Frame
SSMO: Space Science Mission Operations
TIM: Technical Interchange Meeting

## 1. Introduction

The Fermi gamma-ray space telescope is a gamma-ray observatory that is currently operating in low Earth orbit. Launched in 2008, the propulsion system was engineered for initial orbit insertion and final deorbit/reentry. That design scope was expanded when an emergency conjunction detection with KOSMOS 1805 in 2012. In this scenario, the Fermi FDS team performed an off-nominal but successful 1 second thruster collision avoidance burn [1]. Another burn was performed for conjunction avoidance in 2024. In addition, a stuck solar panel in 2018 further restricted operations scheduling on Fermi. Nonetheless, Fermi's 16+ years of service has provided significant scientific contributions to the field of astrophysics and continues to do so [2].

*1.1 Conjunction Risk Motivation*

The increase in spacecraft conjunction alerts from the NASA Conjunction Assessment Risk Analysis (CARA) office have motivated the Fermi team to study a potential orbit change maneuver sequence to protect the observatory. CARA reports of potential conjunctions have increased in recent years with the addition of commercial constellations, introducing further negotiation and discussions on conjunction assessment in the Fermi orbit [3]. Characterizing the orbit raise performance while ensuring NASA Fermi operation and safety is the impetus of this analysis.

*1.2 Summary of Paper*

This paper documents the formulation and analysis framework for Fermi maneuver planning to potentially raise the orbit of the Fermi observatory. The first section details the observatory and force model used to simulate spacecraft dynamics. The second section details the maneuver targeting formulation and an example orbit raise maneuver sequence is presented. Finally, a Monte Carlo analysis capturing how orbit raising performance error can be used to inform a single orbit raising maneuver.

## 2. Maneuver Planning Design

*2.1 Spacecraft Model / Force Model Definition*

For modeling and testing a Fermi maneuver design, Table 1 details the primary properties used at the time of this study. As it is in low Earth orbit, the driving accelerations in order come from Earth two body motion, atmospheric drag, J2 nonspherical modeling, and solar radiation pressure perturbations. Given these design parameters, specific performance characteristics for Fermi maneuver modeling are described in upcoming sections.

**Table 1. Fermi Spacecraft and Force Model Properties**

| | |
|---|---|
| Total Mass | 4356.0 kg |
| Fuel Mass | 353.5 kg |
| Drag Area | 14.18 m$^2$ |
| Solar Radiation Pressure (SRP) Area | 14.18 m$^2$ |
| Fuel Mass | 353.5 kg |
| Nonspherical J2 Model Definition | 70x70 EGM96 |
| Atmospheric Drag Model | Jacchia Roberts, Cannonball |
| SRP Flux | 1370.052 W/m$^2$ |
| SRP Model | Cannonball |
| Tank Pressure | 2005 kPa |
| Tank Temperature | 20 deg C |
| Tank Volume | 0.75 m$^3$ |
| Thruster Thrust | 264 N |
| Thruster Specific Impulse | 221.11 seconds |
| Thrust Scaling Factor | 0.928 |
| ACS Thruster Duty cycle | 92.8% |

*2.2 Propulsion Limitations on Maneuver Performance*

The Fermi propulsion was intended to be used primarily for commissioning and end of life operations, with short burns for orbital debris avoidance [4]. It was not originally designed for use like the studied orbit-raise campaign. As a result of this, a considerable amount of analysis has been required to clear the spacecraft systems for this novel type of use.

From the point of view of the propulsion system and estimation of the amount of hydrazine fuel that would be required, a spacecraft that has been flying as long as Fermi has would be expected to have a well-calibrated propulsion system, with small uncertainties on thrust levels, propellant blow-down curves, etc. Instead, at the start of the orbit-raise maneuver analysis, the only in-flight burn analysis data that was available was from a 1 second collision avoidance burn in 2012. This maneuver had not yet reached steady state, so the data that it provided could not be used to tightly constrain the performance of the propulsion system. Since then, there has been an additional 1 s CA burn in 2024, still insufficient for calibrating the propulsion system to a higher fidelity. Consequently, analysis of the orbit-raise maneuver sequence includes reasonably large margins on key design parameters.

*2.3 Model Assumptions*

The following assumptions have been made in designing a proposed Fermi orbit-raising maneuver sequence. These are biased mainly towards being conservative, i.e. predicting higher propellant consumption than may be seen in practice, since Fermi on-orbit maneuver data is currently limited:
- Required margin of usable fuel at completion of all maneuvers: at least 9 kg (this corresponds to approximately 10% of the fuel used for the largest deorbit burn for end of mission)
- Burns use 10 out of 12 thrusters to be conservative. All burns are purely posigrade or retrograde, as they are each either apogee-raises (AR), apogee-lowerings (AL), perigee-raises (PR) or perigee lowerings (PL)
- Data from report on 2012 2 second CA burn performance:
  – Thrust blowdown curve scaling is shown in Figure. 1. This was used as a reference to model fuel use in maneuver targeting.
  – Specific impulse estimated to be 228.0 seconds for the CA burn. To allow for small reductions as the propulsion system blows down, the value used for the orbit-raising analysis was 221.1 seconds.
- ACS duty cycle 92.8% was a setting chosen based on a worst-case 10-jet value was used for this study. Initial orbital altitude that was used in the initial orbit-raising maneuver design analysis in 2022 was taken as 523 x 539 km. Even with the low A/m value produced by the massive LAT instrument, the orbit has slowly come down during the current solar maximum to a perigee altitude of about 500 km and an apogee altitude of 508 km. For this analysis, the original design orbits in 2022 were used.
- The range of altitudes that is currently deemed suitable for Fermi science is 450-575 km; however, these limits are subject to verification by the science team.

Future work and publications plan on updating these values to the current Fermi performance metrics.

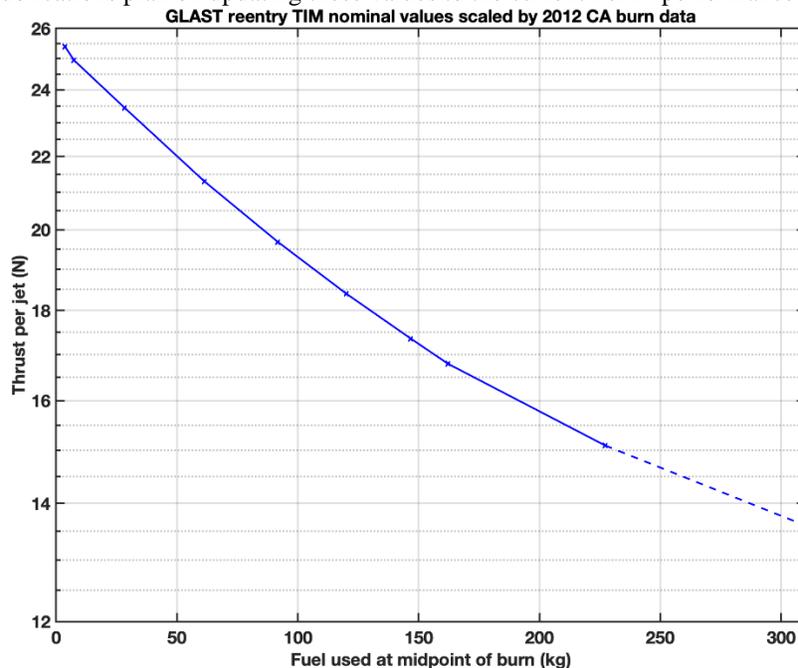

Fig. 1. Propulsion Blowndown Curve (Force vs. Fuel Use) for the Fermi (previously GLAST) Spacecraft

*2.4 Maneuver Design Targeter*

The maneuver design targeter, created in the NASA GMAT software, uses a differential corrector to perform a impulsive targeting problem that dovetails as an initial guess to the finite burn targeting solution. The script is designed for integration into the FDS ground system for future maneuver operations as well as adjusted to Fermi's constrained operational limitations. As such, there is an input for the burn epoch for Fermi maneuver operations scheduling, and the maneuver vector is fixed in the local velocity component direction.

The target state of the maneuver planner is a user selected semi-major axis and eccentricity. A single burn in the Fermi local velocity direction is targeted as the control variable with a user provided initial state and burn epoch. The target semi-major axis and eccentricity tolerances are scaled to reliably allow the single burn to target both parameters and the FDS team's desired orbit. Once the target of the finite burn is achieved, a thrust history file is generated for the operations team. With this approach, the multi-major orbit raising sequence was built for an orbit raise campaign.

*2.5 Multi-Maneuver Orbit Raising Sequence Considerations*

The maneuver sequence starts with smaller burns, with the intention of using these to calibrate the performance of the propulsion system more accurately for the following larger burns that will use more significant propellant. Orbit-raising is designed to allow enough fuel after its completion that the end of mission plan will include a standard deorbit sequence, modified only by the altered final altitude of the operational orbit. A key goal of the orbit-raising sequence is to ensure that sufficient fuel remains after it is completed to still allow a fully controlled reentry. Data produced during the earlier, smaller orbit-raising burns will allow some conservativeness in the later, larger maneuvers to be removed.

*2.6 2022 Example Multi-Maneuver Orbit Change Sequence to a 563 km orbit*

The following maneuver proposal in Table 2. is an example of an orbit raising sequence proposed by the Fermi flight dynamics team which captures an orbit raising sequence and targets an orbit range that would have mitigated earlier conjunction events from CARA.

**Table 2. Maneuver Orbit Raising Sequence From 506 x 526 km, Targeting a 563 x 563 km orbit**

| Burn ID Number | Post Burn Perigee altitude (km) | Post Burn Apogee altitude (km) | Burn Duration | Fuel Used (kg) | Finite Burn ΔV (m/s) | Mean thrust per jet (N) | Burn Type (AR, AL, PR, PL) |
|---|---|---|---|---|---|---|---|
| -- | 506.0 | 526.0 | -- | -- | -- | -- | -- |
| 1 | 508.1 | 526.0 | 10 sec | 1.149 | 0.580 | 26.6 | PR |
| 2 | 512.2 | 526.0 | 20 sec | 2.242 | 1.131 | 26.5 | PR |
| 3 | 526.0 | 526.0 | 1.18 min | 7.521 | 3.800 | 25.2 | PR |
| 4 | 526.0 | 563.0 | 3.29 min | 20.033 | 10.155 | 24.0 | AR |
| 5 | 563.0 | 563.0 | 3.48 min | 19.915 | 10.142 | 22.6 | PR |

As shown in Table 2, initial burns enable thruster calibration and testing using short periapsis raising burns. The third burn circularizes the orbit to an intermediate orbit altitude of 526 km, a region with fewer tracked objects per the CARA analysis [3] to allow the Fermi team to evaluate progress. The last two burns formally raise the orbit to the target of 563 km and has the most maneuver performance risk. To characterize this risk, the following section models a Monte Carlo analysis: using a Fermi GPS position and velocity estimate, it statistically quantifies maneuver performance uncertainty.

**3. Single Burn Monte Carlo Analysis**

*3.1 Problem Statement and Design Assumptions*

To bound Fermi maneuver performance with known hardware performance, a Monte Carlo analysis was made to vary the nominal single maneuver plan with a series of operational performance errors. This immediate step enables the Fermi team to evaluate future test burns. Future work is planned for the entire maneuver campaign such as the one in Table 2.

*3.2 Monte Carlo Methodology*

This section defines the random variable models and Monte Carlo setup; the following section details the Fermi operations process and where errors were inserted to perform each Monte Carlo trial.

**Table 3. Monte Carlo Random Variable Error Properties**

|  | Nominal Value | Mean Bias | 1σ Standard Deviation |
|---|---|---|---|
| Position and Velocity | See Appendix A | 0 km, 0 km/s | See Appendix A |
| Maneuver Magnitude | 0.0064 km/s | 0 km/s | 1.33e-04 km/s + 2.5% of Planned ΔV |
| Maneuver Direction | Velocity Unit Vector | 0 degree | 1.0 degree Elevation Angle |
| Coefficient of Drag (Cd) | 2.1 | 0 | 0.21 |
| Coefficient of Radiation (Cr) | 0.75 | 0 | 0.1 |
| Acceleration Stochastic Errors | 0 km/s² RSS | 0 km/s² | 1e-06 km/s² RSS |

The overall Monte Carlo constants and scenario settings are listed in Appendix A. The chosen initial state in equates to around a 515.7 x 548.8 km orbit and target state equates to approximately a 539.0 x 539.0 km orbit for an epoch ~1.25 days after the burn. Gaussian random variable errors defined with a nominal value, mean error, and standard deviation are all listed in Table 3. The nominal maneuver itself is around 6.4 m/s and performs an orbit circularization by raising perigee. The Monte Carlo consists of 1200 trials based on a target state precision of 10 m, a standard deviation error of the orbit radius of 0.1207 km (calculated based on a smaller 200 trial Monte Carlo run), and a z-statistic of 2.807 for a 99.5% confidence interval. There terms were selected to conservatively bound Fermi maneuver control to 10 meters; enabling position dispersion results to be interpreted on the order of hundreds of meters.

Monte Carlo error sources and parameters are listed in Table 3. The position and velocity error term in the table bounds the uncertainty in the orbit determination estimation process; the best estimated trajectory is taken from the initial cutoff epoch listed in Appendix A and perturbed by the associated covariance, with both the initial state and covariance defined in appendix A. Maneuver magnitude and direction are defined as a sum of a proportional thrust performance error $\epsilon_{scaled}$ and a constant error $\epsilon_{flat}$ modeled on attitude control system thruster contributions. It is summarized in Equation 1-2 below with specific values listed in Table 3:

$$\epsilon_{mag\_err} = \epsilon_{flat} + \epsilon_{scaled} \quad (1)$$

$$\epsilon_{scaled} = 0.025 * \Delta V_{planned} \quad (2)$$

The direction error is split into azimuth and elevation offset from the thrust vector direction and can be seen in Figure 2. The elevation angle models the cone angle offset from the thrust vector and is the driving source of uncertainty. The azimuth angle is defined as a uniform distribution of the full angle range $[0, 2\pi]$.

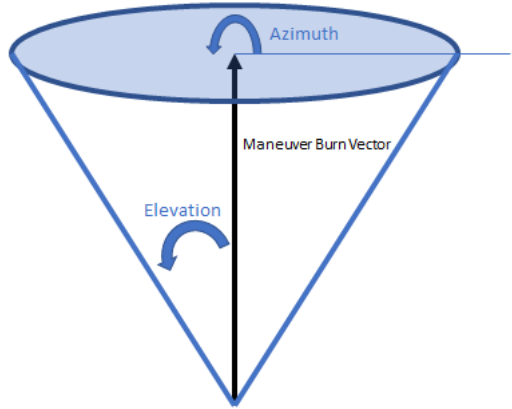

Fig. 2. Maneuver Direction Error Geometry

For force model uncertainty of atmospheric drag and solar radiation pressure, the coefficient of drag (Cd) and coefficient of radiation (Cr) error terms are realized as normally distributed biases off a nominal value defined in Table 3. Finally, stochastic errors were introduced throughout the post maneuver propagation arc to represent any unmodeled accelerations. Scaled in magnitude by the normal distribution defined in Table 3, these stochastic errors were spaced in time and randomized as a random walk scalar noise input process $x(t)$ modeling an array of instantaneous changes in velocity applied every $\Delta t = 12$ hours. For completeness, the relevant equations are listed and are cited from Carpenter and D'Souza [5].

$$x(t + \Delta t) = \Phi(\Delta t)x(t) + w(t) \quad (3)$$

$\Phi(\Delta t)$ is the state transition matrix and $w(t) \sim N(0, S(\Delta t))$ is the noise sample vector with a normal distribution of 0 and standard deviation of $S(\Delta t)$. The Cholesky decomposition $S(t)$ is defined below for the 6x6 state of position and velocity and $I_3$ is a 3x3 identity matrix and $0_3$ is a 3x3 0 matrix:

$$\sqrt[c]{S(t)} = \begin{bmatrix} \frac{\sqrt{3t^3}}{3} I_3 & 0_3 \\ \frac{\sqrt{3t}}{2} I_3 & \frac{\sqrt{t}}{2} I_3 \end{bmatrix} \quad (4)$$

The observed output from the Monte Carlo comes from the final state dispersion, with a primary focus on semi-major axis and eccentricity. A post propagation arc of ~1.25 days from the burn epoch was chosen to ensure sufficient GPS measurements. In addition, the radial-intrack-crosstrack (RIC) state dispersion output was studied to see the spread of Fermi estimate states. The RIC frame is defined as a local frame around the Earth central body; radial is in the radius vector direction, cross-track is in the orbit angular momentum direction, and the in-track vector is the right-hand complement.

*3.2.1  Mission Operations Simulation Setup*

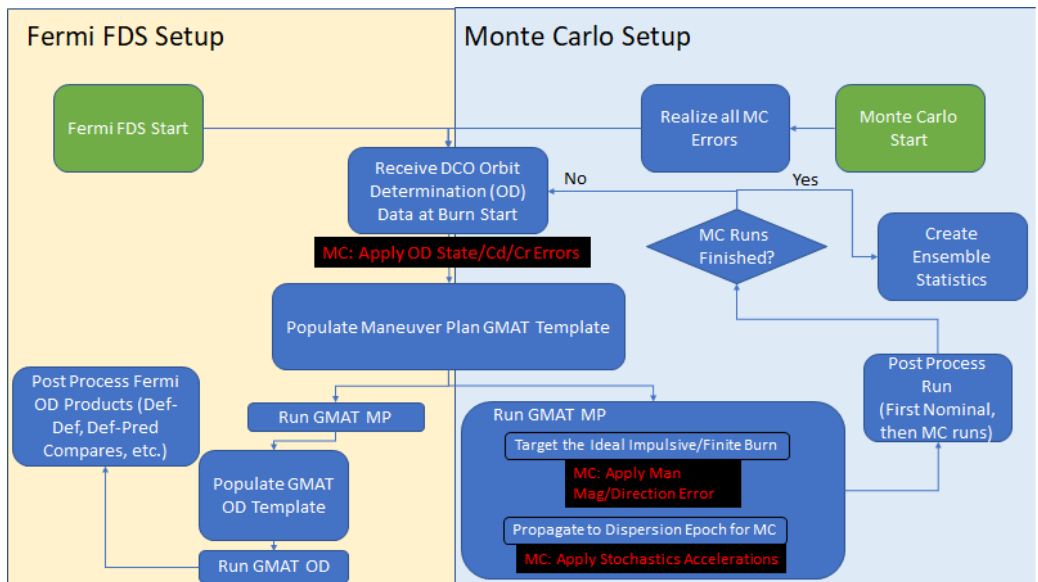

Fig. 3. Fermi Operations Proposed Maneuver Process alongside the Monte Carlo Analysis Process

The original design for NASA Fermi's ground operations did not include maneuvers during routine operations. The proposed NASA Fermi Flight Dynamics System (FDS) ground system design models a maneuver in NASA GMAT software as a follow-on process after performing orbit determination. In the proposed update of FDS, the team can inject an externally designed maneuver into the GMAT orbit determination engine to verify post burn reconstruction. This can be seen on the left side of Fig. 3. The Fermi team in nominal operations generates an orbit determination solution at the data cutoff (DCO) epoch and then runs the GMAT Kalman filter process to generate a definitive and predictive ephemeris state. If desired, a maneuver plan can be executed in a separate GMAT script as an input to the GMAT Kalman filter.

The Monte Carlo process (right side of Fig. 3.) mirrors operations to create a predictive future ephemeris but with realized errors to replicate ground operations. The state uncertainty error is set by the knowledge error of the Kalman filter, so each Monte Carlo trial realizes state error and applies it after the OD state is generated at the data cutoff epoch. The Cd and Cr terms are perturbed at the same step, so errors are applied post orbit determination. Maneuver targeting is then calculated based on that perturbed initial state. Once the maneuver is targeted, magnitude and direction errors are applied for execution. Finally, the post burn propagation arc applies stochastic errors to replicate the impact of momentum unloads and other unmodeled accelerations. The Monte Carlo does not simulate GPS measurements for the newly targeted state in MP. In this analysis, the GMAT MP script propagates the post burn state to generate Monte Carlo results. For published results on filter performance, Vavrina, Newman, Slojkowski, and Carpenter go through the orbit determination process in greater detail regarding conjunction assessment in their paper [6].

*3.3 Results and Discussion*

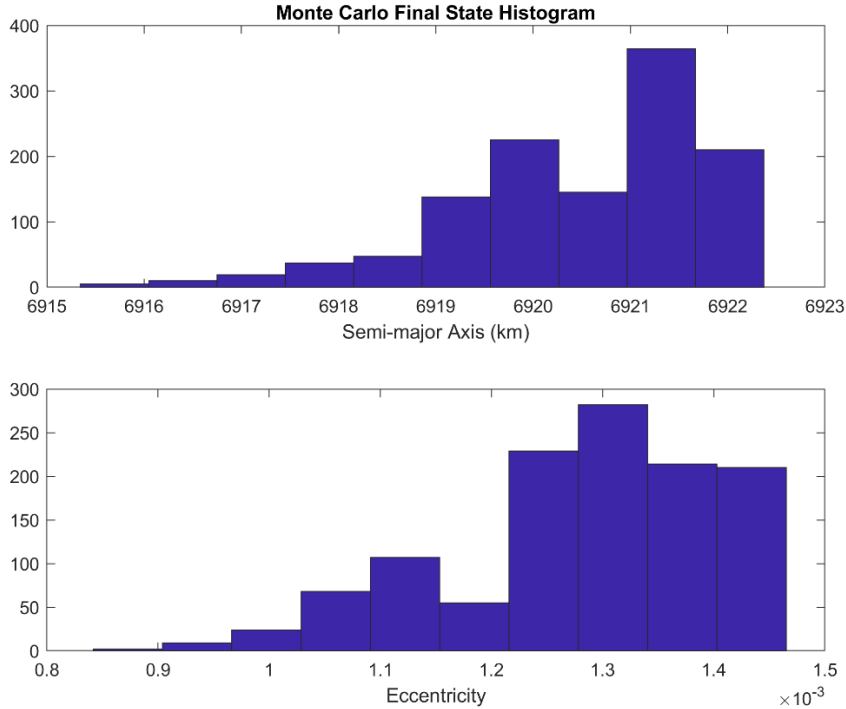

Fig. 4. Monte Carlo Resultant semi-major axis and eccentricity state dispersion

As seen in Figure 4, the mean semi-major axis dispersion is 6920.53 km, with a standard deviation of 1.379 km. The mean eccentricity of the dispersion is 0.0013 and a standard deviation of 0.000122. With a target of 6917 km semi-major axis and 0 eccentricity, this results in a bias of 3.53 km and 0.0013. The output distribution is not Gaussian, so further trials on the models is required to ensure greater statistical significance. One conservative bound on accuracy can be suggested to be the mean bias plus standard deviation; in this case Fermi can reach its target orbit within 5 km semi-major axis and 0.002 eccentricity.

In Appendix B, Figure 5 shows the state dispersions in the Radial-Intrack-Crosstrack (RIC) frame. Looking at the final state output from each of the Monte Carlo cases, the spread is within 0.528 km 1σ in position magnitude and 0.000470 km/s 1σ in velocity magnitude. The direction of the dispersion is primarily in the radial direction in position, with the corresponding error in velocity primarily in the in-track orbit direction.

The major contributor of total dispersion in this study was determined to be maneuver execution magnitude and direction error. The maneuver magnitude error provides the greatest delta in spacecraft acceleration than the other error terms and is primarily in the spacecraft local velocity in-track direction. That impact can be seen with the integrated in-track velocity dispersions in Figure 5. Future work in better modeling propulsion performance with test burns will greatly increase the accuracy of future Fermi orbit change maneuvers.

**4. Conclusions**

This paper presents a maneuver design to orbit raise the Fermi observatory should there again be an increasing risk of spacecraft conjunctions around the 500 km altitude orbit range. The maneuver design supports a mission that currently did not require maneuvers until end-of-life operations. The maneuver design incorporates notional performance limitations with initial checkout burns to slowly test the system, perigee/apogee raising stages with checkout periods, and finally the circularization burn to safely bring Fermi to a new operational orbit. This possible design enables the Fermi operations team to evaluate hardware while maintaining fuel for end-of-life operations.

The Monte Carlo analysis presented in this study demonstrates that a single circularization maneuver could conservatively achieve the target orbit within 5 km semi-major axis and within 0.002 eccentricity. This Monte Carlo analysis shows most dispersions have the greatest velocity vector spread in the in-track direction and the greatest position spread in the radial direction. Cross-track dispersions are significantly smaller than the other components for both velocity and position results. The driving error is maneuver execution error; additional work is necessary for post-test burn calibrations and other maneuver options to confirm maneuver execution error uncertainty. This maneuver design outlines a process for the Fermi FDS team to plan an orbit raise should it become necessary.

**Acknowledgements**

The authors would like to thank Fermi Mission Director Charles McConnell, system engineer lead Julie Halverson, system engineer Rebecca Besser, and Space Science Mission Operations (SSMO) Project Manager Rich Burns for all the support and guidance on this design process.

**Appendix A (Monte Carlo Fermi Initial State, Initial Covariance, and Target State)**

1. Number of Monte Carlo Trials: 1200
2. Orbit Determination Data Cutoff Epoch (UTC): 07 Aug 2022 04:36:09.000
3. Initial Position (MJ2000 frame) km: [-1368.95, -6097.89, 2910.01]
4. Initial Velocity (MJ2000 frame) km/s: [7.46047, -1.35143, 0.685633]
5. Initial Position/Velocity Covariance Uncertainty (MJ2000 frame) km, km/s.
   Format of Row and Columns are: MJ2000 X, Y, Z, Vx, Vy, Vz

[[1.08099e-05, 8.52541e-08, 2.34973e-08, 8.24577e-13, 7.65902e-12, 1.75767e-10],
[8.52541e-08, 9.96553e-07, 1.16061e-07, 6.07633e-13, 1.52655e-14, 7.96717e-15],
[2.34973e-08, 1.16061e-07, 1.42827e-06, 2.84978e-13, 1.25831e-14, 3.57621e-16],
[8.24577e-13, 6.07633e-13, 2.84978e-13, 2.15295e-10, 3.12933e-13, 1.17640e-13],
[7.65901e-12, 1.52655e-14, 1.25831e-14, 3.12933e-13, 6.63691e-11, 8.06399e-13],
[1.75767e-10, 7.96717e-15, 3.57621e-16, 1.17640e-13, 8.06399e-13, 9.21040e-13]]

6. Burn Epoch (UTC): 07 Aug 2022 05:11:09.000
7. Target Epoch (UTC): 08 Aug 2022 11:21:30.809
8. Target Position (MJ2000 frame) km: [-1373.53, -6118.28, 2919.74]
9. Target Velocity (MJ2000 frame) km/s: [7.43993, -1.34464, 0.68228]

**Appendix B (Additional Monte Carlo Results)**

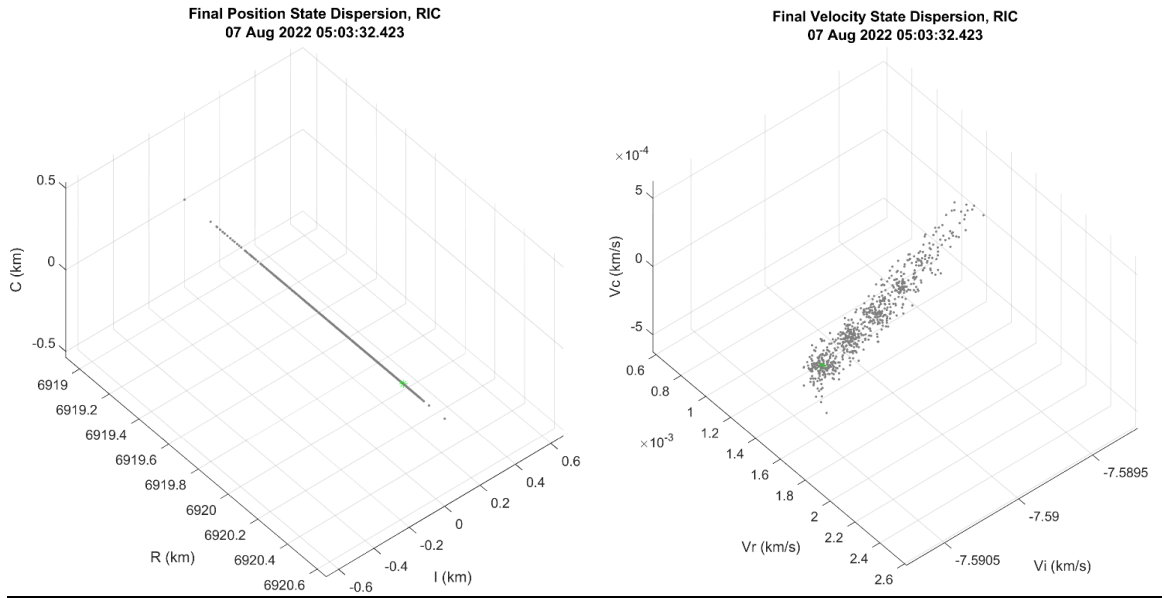

Fig. 5. Final Position/Velocity State Dispersions in the RIC frame. Nominal Value Marked with a Green Asterisk.